\documentclass{ws-p8-50x6-00}

\begin{document}

\title{Sympathetic Cooling of Lithium by Laser-cooled Cesium}

\author{S. Kraft, M. Mudrich, K. Singer,
R. Grimm\thanks{{Permanent address: University of Innsbruck, A-6020 Innsbruck, Austria}},
 A.~Mosk and M.~Weidem\"{u}ller}

\address{Max-Planck-Institut f\"{u}r Kernphysik, 69117
Heidelberg, Germany \\
E-mail: m.weidemueller@mpi-hd.mpg.de}


\maketitle

\abstracts{We present first indications of sympathetic cooling between two
neutral trapped atomic species. Lithium and cesium atoms are simultaneously stored in an optical
dipole trap formed by the focus of a CO$_2$ laser, and allowed to interact for a given
period of time. The temperature of the lithium gas is found to decrease when in thermal contact
with cold cesium. The timescale of thermalization yields an
estimate for the Li-Cs cross-section.
}

Mixtures of atomic gases enrich the field of physics with
ultracold
gases by making many interesting phenomena available for experimental
study. A prime example is the exchange of thermal energy
between different components in the mixture. This
cross-thermalization can be used for determining inter-species
collision cross-sections, and, more importantly, for sympathetic
cooling. The process of sympathetic cooling, where the gas of
interest is cooled through thermalization with a coolant gas, is
especially advantageous if the cooled gas (Li in our case) is hard to cool by other methods,
 while the  coolant gas (Cs) allows for efficient laser cooling.
The Li can only be optically cooled to approximately the Doppler temperature of
 $0.14$ mK, and evaporative cooling of this atom in its stable
 $F=1$ ground state is extremely difficult due to its anomalously
 low scattering length \cite{Schreck0107442}. Cesium, on the other hand,
  allows sub-Doppler temperatures ($< 3\,\mu$K in free
space) to be reached by simple polarization-gradient cooling. This
opens the perspective of reaching a stable Bose-Einstein condensate (BEC) of Li
in the $F=1$ state (as in the experiment by Schreck {\it et al.}
\cite{Schreck0107442})
without evaporative cooling.

\begin{figure}
\centering
\epsfxsize=15pc
\epsfbox{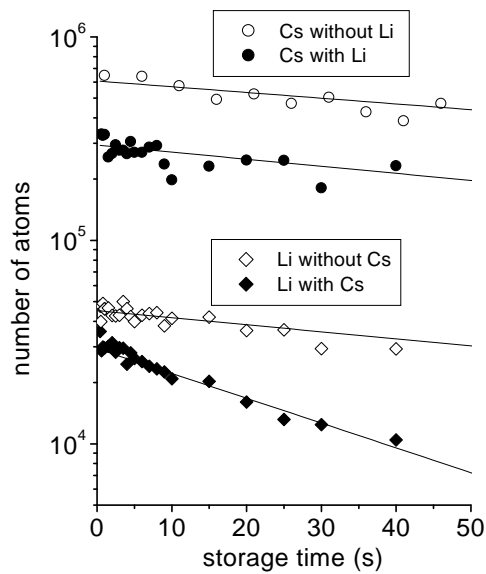}
\caption{Evolution of the number of trapped Li (F=1) and Cs (F=3)
atoms in simultaneous trapping experiments (closed symbols). For
comparison, the open symbols show the evolution of separately
trapped gases. Thin lines: Exponential fits. Time constants are
150 s for Cs with or without Li, 125 s for Li without Cs, and 35 s
for Li trapped simultaneously with Cs.}
\label{fig:evo}       % Give a unique label
\end{figure}

Our apparatus is described in detail in a previous paper \cite{AppPB}.
The thermalization experiments take place in an optical dipole
trap, which is based on a CO$_2$ laser emitting light at 10.6
$\mu$m. As the laser frequency is detuned far below any optical
resonances of the atoms, this trap is referred to as a
quasi-electrostatic trap (QUEST) \cite{Takekoshi}. Our QUEST employs a 140 W laser beam which
is focused to a beam waist of 90 $\mu$m. This yields a
depth of $0.4\,$mK for Li and 1.0 mK for Cs. The axial and
radial oscillation frequencies for Cs atoms are 18 and 800 Hz,
respectively, the corresponding frequencies for Li are higher by a
factor $2.8$ due to the difference in mass and polarizability.

\begin{figure}[t]
\centering
\epsfxsize=20pc
\epsfbox{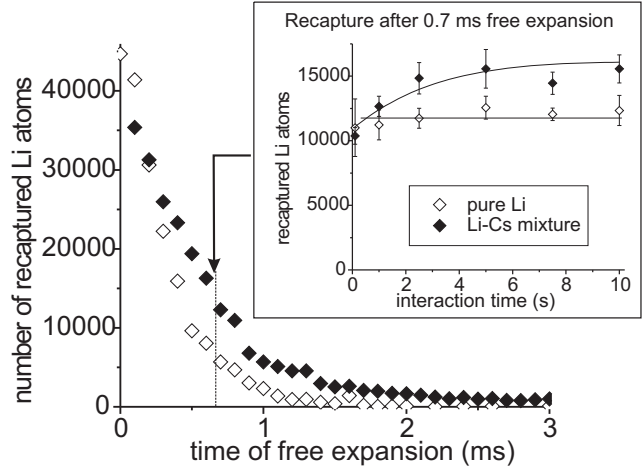}
%\epsfxsize=15pc
%\epsfbox{fig2.eps} % postscript image file name

\caption{\label{fig:recapture} {\it Main graph:} Example of release-recapture measurements. The number of recaptured lithium
atoms is plotted against the free expansion time.
 Open symbols: a Li sample just after loading from the MOT.
Closed symbols: a sympathetically cooled Li sample (interaction time 15 s).
{\it Inset:}
Evolution
of the number of recaptured atoms (free expansion time 0.7 ms). Closed
symbols: number of recaptured Li atoms after sympathetic cooling.
Open symbols: reference measurement with no Cs present. The lines
serve as a guide to the eye.
}
\end{figure}
The atoms are loaded into the QUEST from a combined magneto-optical trap
(MOT) \cite{schloeder99}. We load up to $10^6$ Cs atoms, at a temperature
of $\sim 30\,\mu$K and up
to $10^5$ Li atoms, which have a thermal energy of the order of the trap depth.

In a typical experiment, we trap a mixed sample and let the atoms
interact for a variable time, and then turn off the QUEST. We
subsequently measure the temperature and density of the Cs gas by
absorption imaging (projection of the shadow of the gas cloud on a
camera).
The Li particle number is measured by fluorescence measurements after recapture in
the MOT. The optical density of the Li gas is insufficient to use
absorption imaging, therefore the temperature of the Li cannot be
determined directly.
The evolution of the number of atoms in the trap is shown in
Fig.~\ref{fig:evo}. When stored separately, both Li and Cs
gases decay due to rest gas collisions in $125$ and 150~s,
respectively. There is no indication of evaporation in either
case: Cs does not evaporate as its temperature is far below the
trap depth. Li evaporation is energetically possible but the cross
section for Li-Li collisions is so small that the characteristic
timescale is $> 1000$~s.

When the gases are stored simultaneously, the Li atom number
shows a non-exponential initial behavior followed by an approximately
exponential decay with a characteristic time of 35~s, which is much faster
than the 125~s rest gas induced decay. We interpret this as
evaporation due to elastic collisions with the Cs.
From the evolution of the particle number we
can determine inelastic and elastic heteronuclear collision
rates \cite{AppPB}. However, the clearest signature of sympathetic
cooling, reduction of the temperature of the cooled gas, cannot be
obtained this way.

To determine the lithium temperature change due to sympathetic
cooling, we have performed release-and-recapture measurements, in a
manner analogous to experiments by O'Hara {\it et al.} \cite{OHara00}.
In this type of measurement, the QUEST is turned off and the
atoms are released in free space for a short time.
After a free expansion time of up to 2 ms the QUEST is turned on
again. The fraction of Li atoms that were recaptured in the QUEST
is a measure of the Li temperature, since slow atoms take a longer
time to leave the trapping region. A more quantitative analysis of these release-recapture measurements
is
being performed and will be published elsewhere.

In figure \ref{fig:recapture} release-recapture thermometry is
demonstrated on two different Li samples: one which has been
loaded from the MOT and stored in the QUEST for a short time
without  Cs, and one which has been
stored together with a cold Cs sample for 15 s. The figure clearly
shows that in the latter case more atoms are recaptured at long times,
i.e., a larger number of slow atoms is present.
 This indicates that the
lithium is cooled sympathetically by the laser cooled cesium atoms.

The timescale of thermalization can be estimated from the inset in Fig.~\ref{fig:recapture},
 where a single point from the
release-recapture curve was measured for different interaction
times. A thermalization
timescale of roughly 3 seconds can be estimated. From a rate equation
model \cite{AppPB}
one estimates elastic scattering cross-section of
$\sigma_{\rm CsLi}\sim 3 \times 10^{-12}\, {\rm cm}^{2}$ which is
in satisfactory agreement with the earlier estimate we obtained from
evaporation measurements \cite{AppPB}. Note that the {\em absolute} number of
slow atoms increases during thermalization, indicating an
increasing phase space density.

Further efforts will concentrate on lowering the
temperature of the coolant gas and increasing the number of trapped Li
atoms,
thereby increasing the degeneracy parameter. By repeated pulsed
optical cooling one can  extract the entropy that the Li
transfers to the Cs, thereby providing a truly lossless cooling
mechanism towards BEC of Li.
For a sample of $10^6$ Li atoms, the BEC
transition temperature is $\sim 3\, \mu$K in our trap.

%\section*{Acknowledgments}
\vspace{1ex}
{\footnotesize This work has been supported in part by the Deutsche
Forschungsgemeinschaft. A.M. is supported by a Marie-Curie
fellowship from the European Community programme IHP under
contract number CT-1999-00316. We are indebted to D.~Schwalm for
encouragement and support.}

%\section*{Appendix}


\begin{thebibliography}{9}
\bibitem{Schreck0107442} F. Schreck {\it et al.}, Phys. Rev. Lett. \textbf{87}, 080403
(2001).

\bibitem{AppPB} A. Mosk {\it et al.},
submitted to Appl. Phys. B; arXiv:physics/0107075

\bibitem{Takekoshi} T. Takekoshi and R. J. Knize, Opt. Lett. {\bf
21}, 77 (1996).


\bibitem{schloeder99}
U.~Schl\"oder \textit{et al.}, Eur.\ Phys.\ J.\ D \textbf{7}, 331
(1999).

\bibitem{OHara00}
K. M. O'Hara \textit{et al.},
%, M. E. Gehm, S. R. Granade, S. Bali, and J. E. Thomas,
% ``Stable, Strongly Attractive, Two-State Mixture of Lithium Fermions in an Optical
%Trap'',
Phys.\ Rev.\ Lett.\ {\bf 85}, 2092 (2000).

\end{thebibliography}
\end{document}